\documentstyle[aps,multicol,epsfig]{revtex}
\draft
\def\PRL{Phys. Rev. Lett.}
\def\vep{\varepsilon}

\def\matr#1{\underline{\underline{{\bbox{#1}}}}}
\def\lz{\ell_{\parallel}}
\def\lp{\ell_{\bot}}
\def\bea{\begin{eqnarray}}
\def\eea{\end{eqnarray}}

\begin{document}
\title{Slow stress relaxation in randomly disordered nematic elastomers and gels}
\author{S.M. Clarke and E.M. Terentjev}
\address{Cavendish Laboratory, University of Cambridge, 
Madingley Road, Cambridge CB3 0HE, U.K.}
\date{\today}
\maketitle
\begin{abstract}
Randomly disordered (polydomain) liquid crystalline elastomers align under stress. 
We study the dynamics of stress relaxation before, during and after 
the Polydomain-Monodomain (P-M) transition. The results for different materials
show the universal ultra-slow logarithmic behaviour, especially pronounced in the
region of the transition. The data is approximated very well by an equation
$\sigma(t) \sim \sigma_{eq}(\vep) + A/(1+ \alpha \log \, t)$.
We propose a theoretical model based on the concept of
cooperative mechanical resistance for the re-orientation of each domain, attempting 
to follow the soft-deformation pathway. The exact model solution can be approximated
by compact analytical expressions valid at short and at long times of relaxation, 
with two model parameters determined from the data.
\end{abstract} 

\vspace{0.5cm}
\begin{multicols}{2}

Liquid crystalline ordering in a confined, randomly quenched geometry has 
been the subject of considerable research in recent years. 
Non-aligned polymer-stabilised or dispersed liquid crystals and, 
in particular, the nematics in aerogels or other porous media 
\cite{clark1,garland,copic} have been studied with an eye on the
universal properties of systems with weak random fields. The basic 
scaling observation, that at long length scales the static random fluctuations
dominate over the dynamic thermal fluctuations, has been verified by rigorous
theoretical analysis and experimental studies of correlations and susceptibilities 
in a wide variety of systems \cite{dozenko}. The critical slowing down of all
relaxation processes in systems with a random, glass-like order
is also a well-established universal phenomenon \cite{ising,bouchaud}, with a 
characteristic stretched-exponential or power-law behaviour as opposed to a normal
exponential relaxation of overdamped ordered systems. Accordingly, it
has not been surprising to find slow relaxation modes in randomly confined 
nematic liquid crystals \cite{copic,goldburg,clark2}. However, although the 
concept of a random orientational effect imposed by the large amount of inner 
surfaces seems intuitively correct, the direct application
of continuum random-field models to systems with strong anchoring on sub-micron 
length scales is more difficult.

Nematic elastomers and gels have been a focus of extensive recent research for a
number of different, but equally compelling reasons. In these systems the liquid crystal 
ordering and director are coupled to the mechanical degrees of freedom --
stresses and strains of the underlying polymer network.  As a result, local elastic 
torques are unbalanced and the non-symmetric Cosserat-like elasticity leads to a 
number of unique physical effects such as soft elasticity and mechanically-driven 
orientational switching (see the review \cite{review} for details). It 
has been argued that the network crosslinks could act as local sources of quenched 
disorder and the nematic gel is analogous to the spin glass with random 
magnetic anisotropy \cite{fridrikh}. The application of external stress to 
(polydomain) nematic networks disordered on large scales 
results in a critical transition into the aligned state with an increasing degree 
of long-range order. This polydomain-monodomain (P-M) transition, analogous to the 
spin-glass alignment by magnetic field, has been studied experimentally in some detail 
\cite{stu}, confirming that the alignment proceeds via the reorientation of correlated 
regions (`domains') rather than the growth of the favoured ones. Because the
polymer chains within each domain are anisotropic, the rotating director causes 
shape changes, Fig.1, and thus the materials can accommodate the external
deformation without significant stress response. In this Letter we study the dynamics 
of this transition by increasing the extensional strain $\vep$ in controlled small 
steps and monitoring the relaxation of stress $\sigma(t)$. 

\begin{figure}[h]
\centerline{ \epsfxsize=7cm \epsfbox{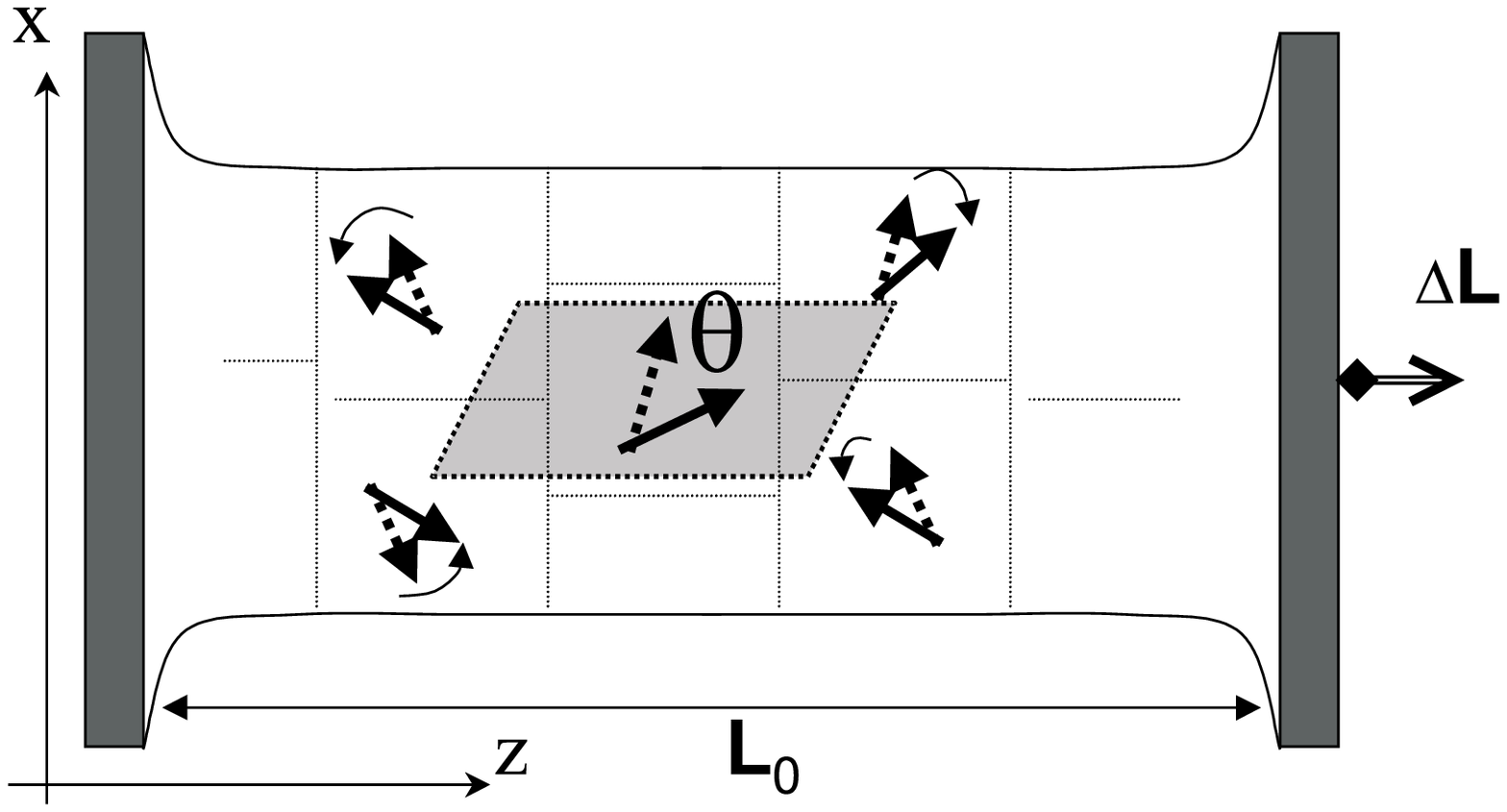}}   
\label{fig1}     
\end{figure} 
\vspace{-0.4cm}
\parbox[t]{8cm}{\scriptsize FIG.1 \ \ Schematic illustration 
of a stretched polydomain elastomer. 
As each correlated domain rotates its director towards the extension axis, its shape 
is changing. The shear strain associated with the rotation $\theta$ 
(indicated by shading of a single domain) needs to be 
compatible with the surrounding elastic medium.}   
\vspace{0.3cm}

We find this relaxation to be extremely slow, increasingly so in the vicinity of 
critical stress for the P-M transition, and following the logarithmic law over several
decades of time. Similar logarithmic decays has been observed in other systems, as different 
as nematics in silica gels 
\cite{goldburg} and avalanches at the angle of repose \cite{nagel}, 
and are reminiscent of the $1/f$ noise kinetics \cite{weissman}.
We propose a theoretical model for a rate constant vanishing with an essential
singularity due to the cooperative mechanical barriers for each domain's rotation.
The resulting kinetic equation, having a form $\dot{\theta}$=$ 
- m \theta^3 e^{-u/\theta}$, gives a solution which at short times 
resembles the power-law $t^{-1/2}$ and at long times
can be approximately interpolated as $1/(1+2\log \, t/t^*)$, with the crossover time
$t^*=(m \, u^2)^{-1}$. Both regimes correspond well to the experimental results. 
We expect this 
behaviour to be universal across the randomly disordered systems where, as in nematic
elastomers and gels, the relevant order parameter is coupled to the elastic modes: 
the mechanical compatibility requires the cooperative nature of elastic barriers. 

The materials used in this experiment are three different types of side-chain liquid 
crystalline polymers: polysiloxanes \cite{kupfer} crosslinked via rather flexible 
tri-functional groups, polyacrylates crosslinked by bi-functional 
chemical groups \cite{mitchell} and by $\gamma$-radiation \cite{talroze}. All materials
had slightly different crosslinking density, around 5-10\%. The backbone 
anisotropy is very different for siloxane and acrylate chains, the anisotropy of
chain radii of gyration being $R_\parallel/R_\bot \approx 1.6$ and $1.06$, respectively
(for Gaussian polymers $\langle R^2_{\parallel, \bot}\rangle
=\frac{1}{3}\ell_{\parallel, \bot}L$
with $L$ the chain contour length). 
The glass transition temperature for polysiloxane 
elastomer was around -5C (with nematic-isotropic transition $T_{ni}\sim$ 42C), while the 
polyacrylates become glasses at $T_g \sim$ 50C (with $T_{ni}\sim$ 110C) \cite{foot}. 
Hence the experiment was carried at 30C for polysiloxane elastomer and at around 90C 
for both polyacrylates. In spite of
all these differences, the results obtained are distinctly similar and in this Letter 
we concentrate mostly on the polysiloxane system, which has more dramatic quantitative 
effects due to the higher chain anisotropy. The experimental approach is straightforward.
Strips of polydomain nematic rubber ($\sim$15{\sf x}5{\sf x}0.3 mm) were suspended on
a stress gauge and extended in controlled strain fashion in a box which was thermostatically
controlled. We
applied a consecutive fixed-step extensions of $\sim$0.5 mm every 24 hours (thus
providing an effective strain rate of $\dot{\vep}\sim$3$\cdot 10^{-7}\hbox{s}^{-1}$). 
The data for variation of response force with time has been collected 
and then converted to the nominal stress $\sigma$ 
 (and the extension -- to engineering strain $\vep=\Delta L/L_0$). 
The question of temperature control required a serious attention. 
The accurate measurement of very small changes in stress over large 
time intervals demanded the experimental error to be brough down to 0.05\% in stress
and 0.1$^{\sf o}$ in temperature. The partially and fully aligned nematic elastomers
are very thermally sensitive. We, therefore, included an additional correction to account
for this effect. 

The stress-strain dependence of polysiloxane nematic rubber, going through the P-M 
transition is shown in Fig.2. The sets of data 
points shown are collected at fixed intervals of 20s, 100s, 500s and 24 hours
after the strain increment from the previous-day value. 
One clearly sees the major trend of the P-M 
transition, exhibited at all times: When the threshold stress is reached, the correlated
nematic domains begin to rotate towards the direction of extension. Because the 
average polymer backbone is prolate along the local nematic director {\bf n}, the 
aligning of average {\bf n} results in the effective lengthening of the whole sample, 
thus accommodating the imposed strain and creating the stress plateau over a large 
interval of deformations. The strain at the end of the equilibrium stress plateau is 
determined by the anisotropy of chain shape $R_\parallel/R_\bot$. 
However, the question of what is the equilibrium stress at each value
of $\vep$ requires a detailed analysis of its kinetics. The inset in Fig.2 
shows the analogous plot for a chemically crosslinked polyacrylate, with a much 
smaller backbone anisotropy and the stress plateau region.

\begin{figure}[h]
\centerline{ \epsfxsize=7cm \epsfbox{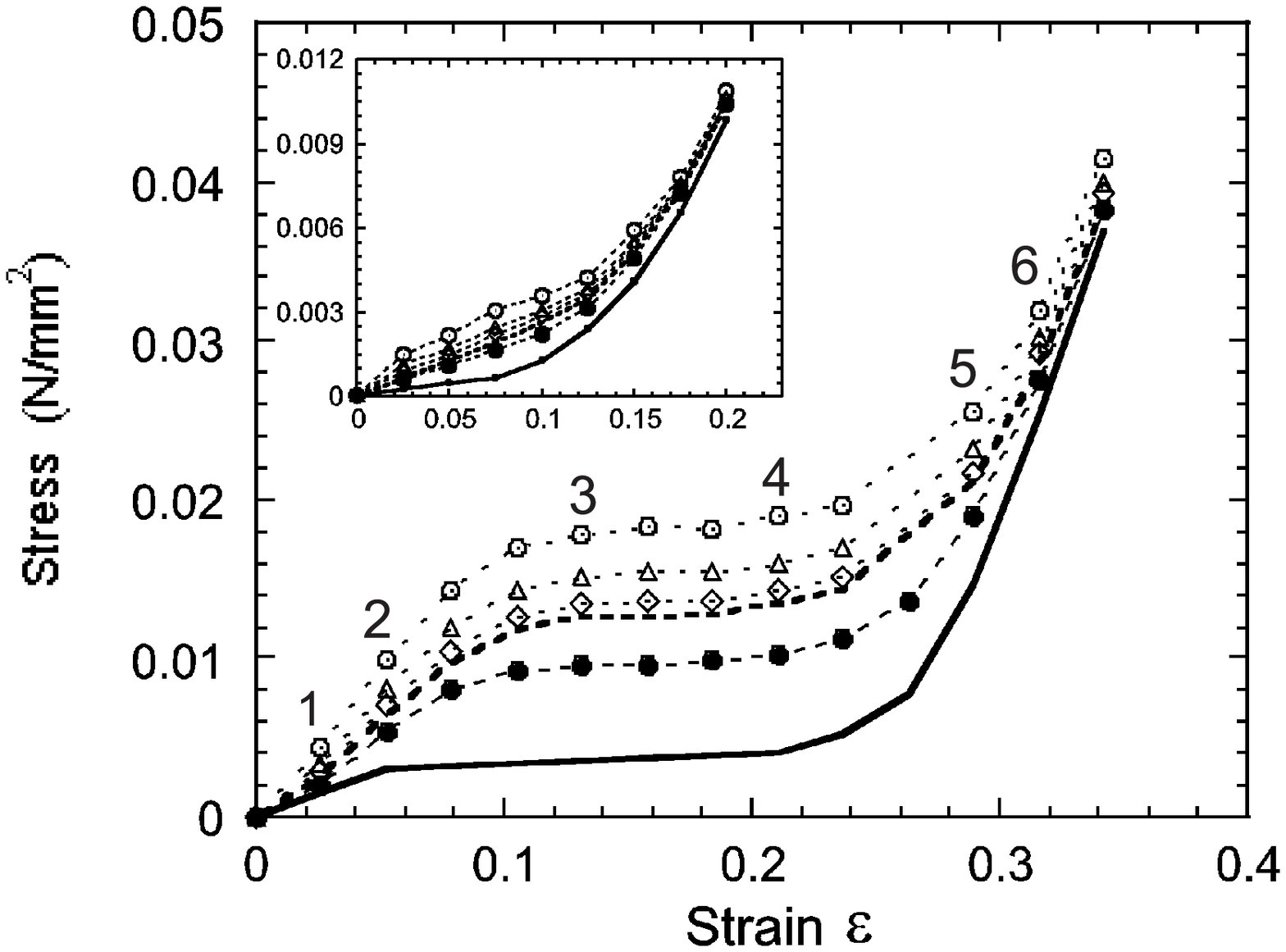}}   
\label{fig2}     
\end{figure} 
\vspace{-0.5cm}
\parbox[t]{8cm}{\scriptsize FIG.2 \ \ Experimental values of the nominal 
stress in polysiloxane nematic rubber, as observed 
at 20s ($\odot$), 100s ($\triangle$), 500s ($\diamondsuit$) and 24 hours ($\bullet$) 
after the strain increment. The numbered points correspond to the data sets used in Fig.3.
The thick dashed line is the saturation level of `apparent power-law'; the solid 
line shows the true equilibrium values $\sigma_{eq}(\vep)$ the stress would achieve, 
if given time for inverse-logarithmic decay to saturate [see Eq.(\ref{inv}) and Fig.4].
 The inset shows the same plot for chemically crosslinked polyacrylate.}   
\begin{figure}[h]
\centerline{ \epsfxsize=7cm \epsfbox{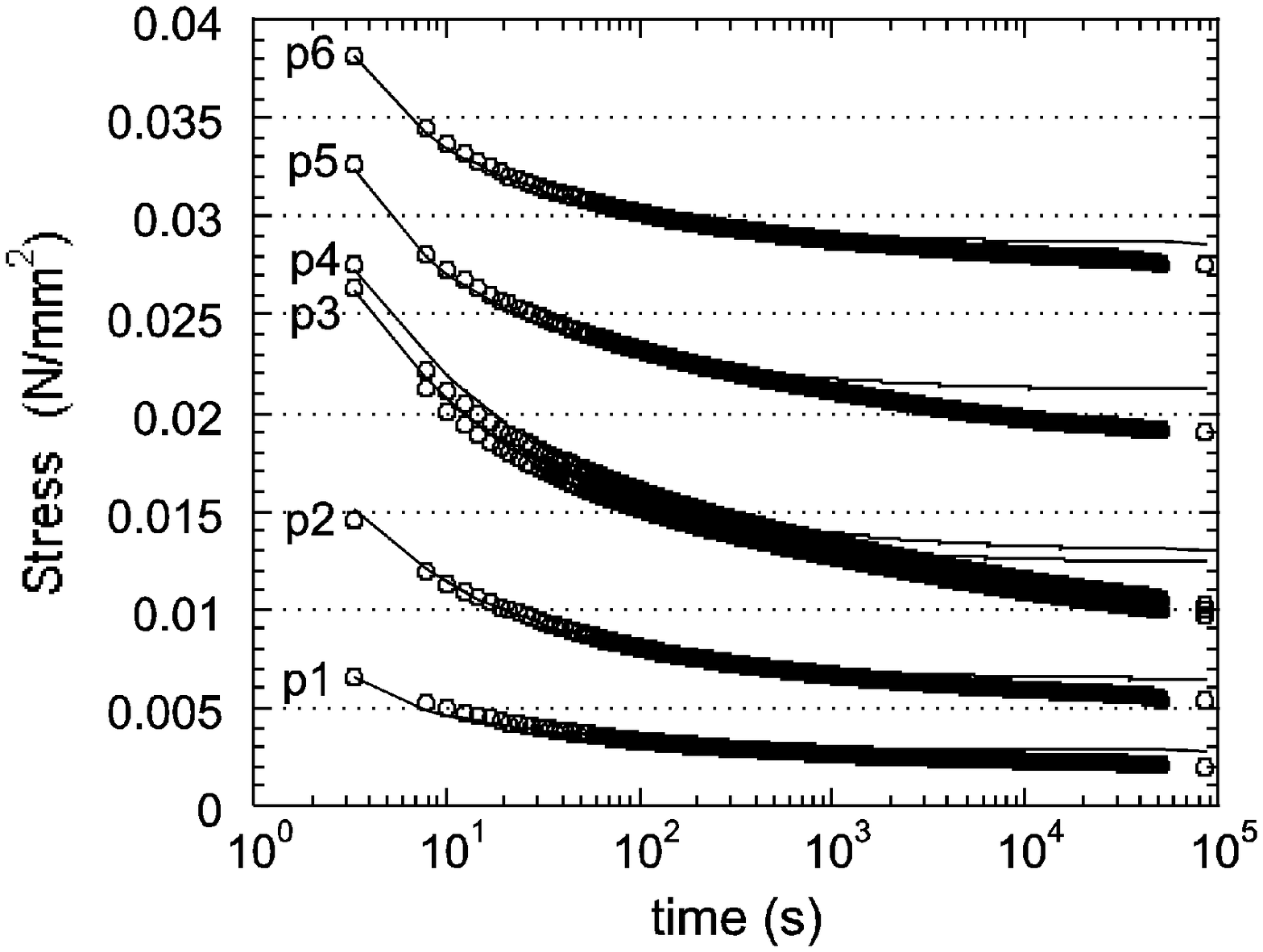}}   
\label{fig3}     
\end{figure} 
\vspace{-0.5cm}
\parbox[t]{8cm}{\scriptsize FIG.3 \ \ Six selected relaxation plots $\sigma(t)$ for
points, labelled from 1 to 6 in Fig.2, in polysiloxane sample. 
Note, that the plots for points p3 ($\vep=0.13$) 
and p4 ($\vep=0.21$) are hardly distinguishable, even though they are at each end of
 the stress plateau. The slow (logarithmic) relaxation at long
times is evident, with the slope significantly increasing as the P-M transition is
being approached from both sides. Thin solid lines show the `apparent power law' model,
Eq.(\ref{inv}), which fits only the short-time region of each plot.}   
\vspace{0.3cm}

Fig.3 shows the variation of stress with time after the strain increment for several
selected points in Fig.2. The logarithmic scale contracts the long-time region
and enhances the effect of decreasing $\sigma(t)$. One immediately recognises the fact 
that, especially for the points on the stress plateau ({\it i.e.} during the P-M 
transition), the equilibrium stress has not been achieved even after the 24-hour 
interval. Fig.4 gives an example of data analysis. 
The relaxation of stress at relatively short times
accurately follows a power-law with a seemingly universal exponent -$\frac{1}{2}$. 
This variation has been reproduced in all three different materials (studied at
different temperatures), for all points on the stress-strain curve of Fig.2. 
At longer times a logarithmic decay was evident. Again, this has been a feature
for all materials and at all strains, however, the slope of the logarithmic decay
was very different, dramatically increasing for the materials with higher chain
anisotropy and towards the P-M transition stress plateau. In fact, the logarithmic decay
was hard to detect for the first and last points in Fig.2 for highly 
anisotropic polysiloxane nematic rubber and, equally, for the chemically crosslinked
polyacrylate (which is known to have a very small backbone anisotropy). Finally, the
elastomer
samples at large deformations (of 10-30\% in our study) always develop a distinct necking
near the rigid clamps (which do not allow the reduction of sample width required
by incompressibility). As a result, even at deformations long past the stress plateau
and the samples well-ordered in the middle, there are still polydomain regions near
the clamps are in an incomplete state of
alignment (this is clearly seen by the eye and under the
microscope). Similarly, well before the P-M transition is reached, there are
internal processes associated with domain wall localisation \cite{fridrikh}, 
also contributing to slow cooperative relaxation. 
Therefore, it is attractive to attribute the residual logarithmic decay away from the 
P-M transition to such artefacts of non-uniformity and conclude that
the main stress relaxation occurs during the P-M transition and is due to the 
rotation of correlated nematic domains.

\begin{figure}[h]
\centerline{ \epsfxsize=7cm \epsfbox{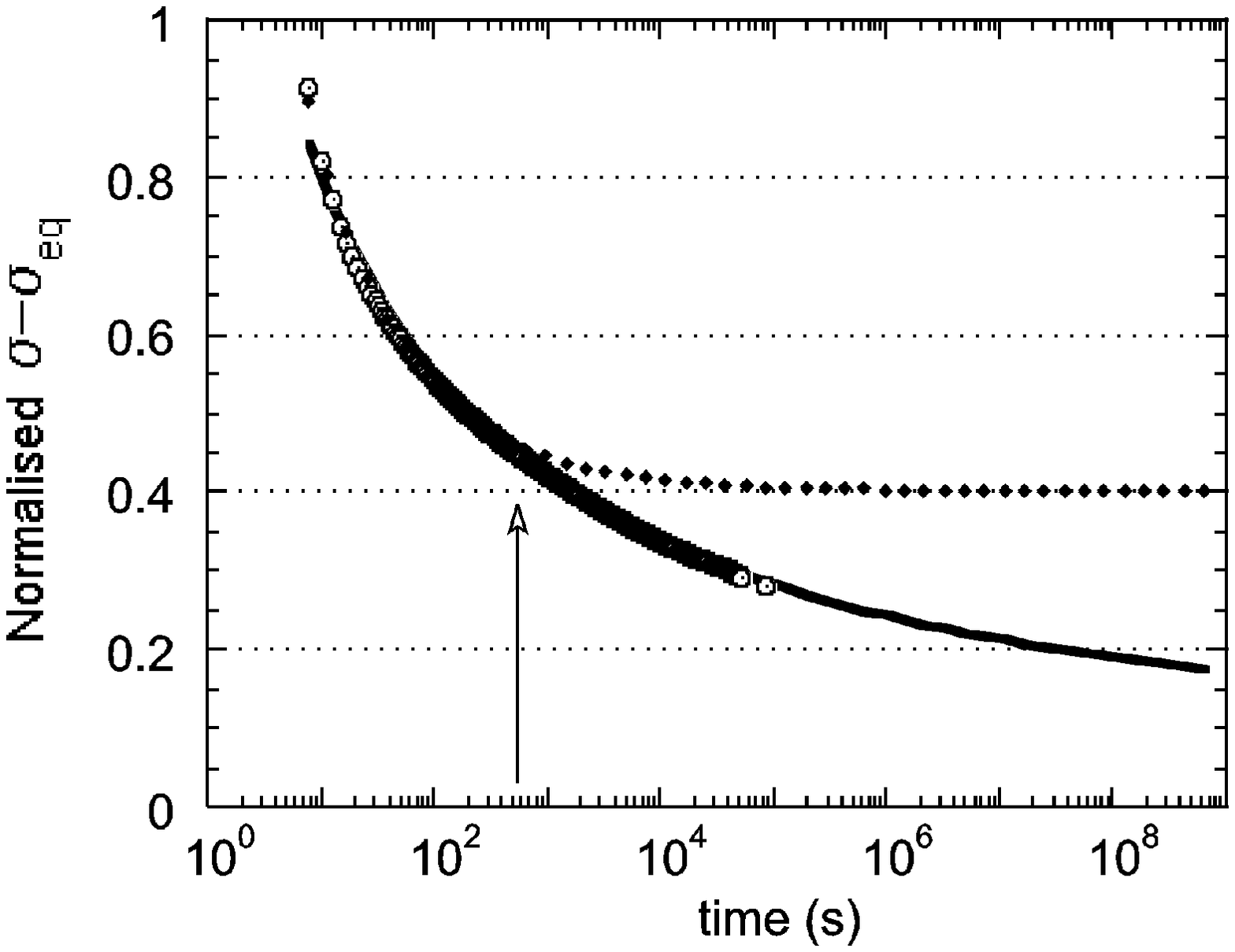}}   
\label{fig4}     
\end{figure} 
\vspace{-0.5cm}
\parbox[t]{8cm}{\scriptsize FIG.4 \ \  
Analysis of relaxation for the point $p4$ ($\odot$) where the stress has been
normalised for convenience to vary between 1 and 0. 
The `apparent power law' model with the exponent $-\frac{1}{2}$
(dots) fits the data well until the crossover time $t^*$ is reached.   
A good overall fit of the stress $\sigma(t)$ is given by the inverse-logarithm
function $\sigma=\sigma_{eq}+A/(1+ \alpha \log \, t)$, the solid line. }  
 \vspace{0.25cm}

To understand the results, one needs to recall the basic concepts of soft elasticity 
in nematic gels (or any other Cosserat-like elastic media with an independent 
Goldstone degree of freedom); the reader should refer to \cite{review} for details. 
Considering a region of locally uniform director (a `domain'), the rubber-elastic
energy can be reduced to zero if the director rotation $\theta$ 
and the Cauchy strain tensor $\matr{\lambda}$ are related by the equation 
$\matr{\lambda}=\matr{\ell}(\theta)^{1/2}\matr{\ell}(0)^{-1/2} $, where 
the uniaxially anisotropic polymer step-length matrix is given by $\ell_{ij}=
\lp \delta_{ij} +
(\lz-\lp)n_in_j$. For a plane stretching and director rotation as in Fig.1, when 
the deformation $\vep =
\lambda_{zz}-1$ is applied, the elastic response can be reduced if the director
rotates by $\theta \approx \sqrt{2\lz/(\lz-\lp)}\vep^{1/2}$. To comply with 
the soft-elasticity pathway, this rotation requires an associated shear 
$\vep_{zx}\approx \frac{1}{2}(\lz/\lp -1) \sin \, 2 \theta$ \cite{review}. 
Therefore, a given domain subjected to an external extension $\vep$
finds itself out of elastic equilibrium, with a director misaligned by $\theta$ 
and the effective energy density
$\Delta f \approx \frac{1}{2} \mu (\lz/\lp -1)^2 \theta^4$ at small $\theta$, 
where $\mu \approx n_{\sf x} k_{\scriptsize B}T \sim 10^5 \hbox{J/m}^3$ is the 
rubber modulus. The relaxation of stress then proceeds via the director rotation
towards its soft-elasticity equilibrium, $\Delta \sigma(t) \sim 
\mu (\lz/\lp -1) \theta (t)$, while the dynamics of $\theta$ is controlled by the
standard model-A equation $\dot{\theta} = -\widetilde{m} \, \theta^3$. Note the
cubic force, which is the direct consequence of soft elasticity and leads to
the power-law decay $\theta = (2 \widetilde{m} \, t + \frac{1}{\theta_0^2})^{-1/2}$, 
where the initial misalignment $\theta_0$ is the function of strain step. 
This power law is indeed observed at early stages of relaxation for all 
materials and deformations studied. 
However, the long-time logarithmic behaviour requires a more delicate analysis.

So far we have discussed the soft response of an individual domain. Surrounded by
its neighbours, each of which has its own director and a different set of 
soft strains, any given domain will face an elastic barrier for its relaxation. 
In an extreme situation when all neighbouring domains are already aligned and
cannot find any soft pathway to accommodate the external strain, 
the shear deformation of the given domain
will not be allowed by mechanical compatibility and no relaxation would occur 
\cite{rubber}. Therefore, for each individual domain to relax its local stress, one
requires a cooperation of its neighbours, providing a `gap' for the required set 
of strains. We shall approximate this effect by estimating the effective rate 
constant $\widetilde{m}$ of escape over the average non-soft barrier.
Let us first introduce the {\it mean} angle $\langle \theta(t) \rangle$, the 
average of misalignment angles of all different domains and thus the measure of 
non-relaxed part of the stress. As $\langle \theta \rangle$ decays to zero 
after each strain increment, the total barrier in the system increases as 
$\langle \theta_0 \rangle - \langle \theta(t) \rangle$ with the initial
condition $\langle \theta_0 \rangle$ proportional to the strain step. 
If we assume that this energy is evenly distributed between all non-relaxed
domains (the number of which is $\propto \langle \theta \rangle$), the 
effective escape rate becomes $\widetilde{m} \sim \exp [-\beta v 
\frac{\langle \theta_0 \rangle - \langle \theta \rangle}{\langle \theta \rangle}]$,
leading to the dynamical equation
\begin{equation}
\frac{d}{dt}\langle \theta \rangle = -m \, e^{-u/\langle \theta \rangle} \, 
\langle \theta \rangle^3   
\label{kinetic}
\end{equation}
(with a small re-arranging of constants). We expect the parameter $u$ to be proportional
to the strain step and to the overall strain $\vep$, which is a practical measure of 
the P-M transition, Fig.2.  Both $m$ and $u$ should also be proportional to the chain 
anisotropy $(\lz/\lp -1)$. Eq. (\ref{kinetic}) integrates for times between zero and $t$ 
and gives
\bea
t &=& \frac{1}{m \, u^2}\left( \frac{u-\theta}{\theta}e^{u/\theta} - 
\frac{u-\theta_0}{\theta_0}e^{u/\theta_0} \right) \ ; \nonumber \\
\theta &=& \frac{u}{1+{\sf ProductLog} \left[ \frac{m \, u^2}{e}t -
(1-\frac{u}{\theta_0})e^{-1+u/\theta_0} \right] }
\label{sol}
\eea
with the direct solution $\theta(t)$ expressed via the special function 
${\sf ProductLog}[z]$, 
a solution for $w$ of $z=w \, e^w$ (we have omitted the $\langle .. \rangle$ 
notation for 
the mean angle $\theta$). The constant in its argument is $\approx 1/e$ at expected
$u /\theta_0 \ll 1$. 
It is straightforward to plot the solution (see Fig.5) and verify that
the essential singularity in the rate constant at $\theta \rightarrow 0$, 
Eq. (\ref{kinetic}), results in a critical slowing down of the long-time decay. 

\begin{figure}[h]
\centerline{ \epsfxsize=7cm \epsfbox{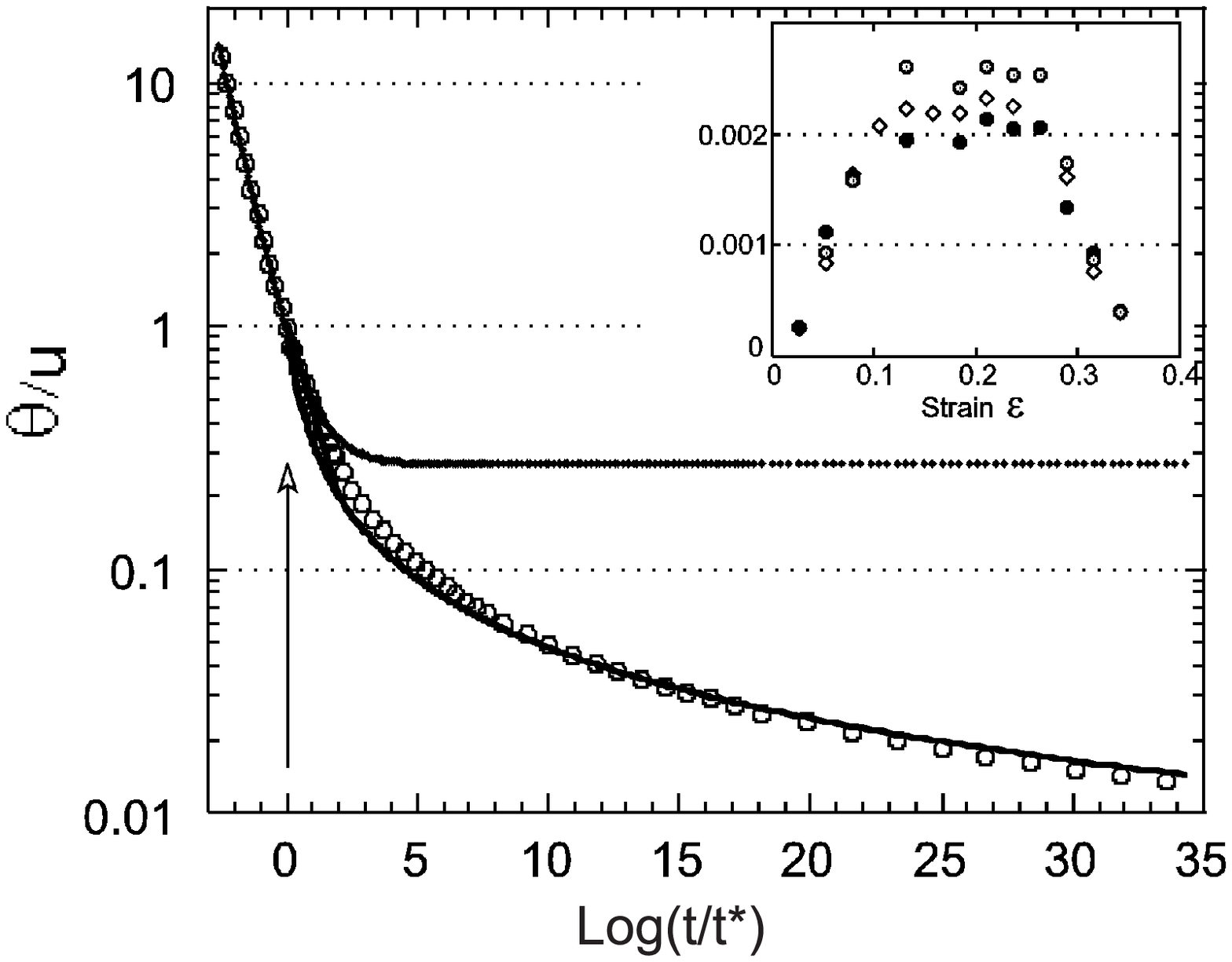}}   
\label{fig5}     
\end{figure} 
\vspace{-0.35cm}
\parbox[t]{8cm}{\scriptsize FIG.5 \ \ The log-log plot of scaled solution $\theta/u$ 
against scaled time $t/t^*=m u^2  t$, as suggested by Eq.(\ref{sol}), 
circles. The arrow shows the position of crossover time with two approximate regimes, 
given by Eqs.(\ref{inv}), shown by dots (`apparent power law') and solid line 
(inverse logarithm). The inset shows the model activation parameter $u$ vs. $\vep$,
determined in three ways described in the text: (i) - $\diamondsuit$, (ii) - $\odot$ and 
(iii) - $\bullet$.}  
\vspace{0.3cm}

\noindent There are several qualitative conclusions to be drawn from the 
Eqs.(\ref{kinetic})-(\ref{sol}). 
At short times, when $\theta$ is not small, the exponential in the 
decay rate is irrelevant and we recover the power-law behaviour 
$\theta \sim t^{-1/2}$. 
This `apparent power law' becomes invalid at a
crossover time $t^* \approx (m \, u^2)^{-1}$ (when $\theta \sim u$), 
after which the inverse-logarithm 
function gives an excellent interpolation of the exact solution (\ref{sol}):
\bea
\theta & \approx & \frac{1}{(2m \, t)^{1/2}} + 0.27 \, u \qquad{\rm for}\ \ 
\frac{1}{m\theta_0^2} \ll t \ll t^*  \nonumber   \\
\theta & \approx & \frac{u}{1+2 \, \log (m \, u^2 \, t)}\qquad \ \, {\rm for}\ \ t \gg t^*
\label{inv}
\eea
The corresponding stress relaxation behaviour, 
$\sigma(t)=\sigma_{eq}+{\rm const} \cdot \theta(t)$
closely reproduces the data, presented in Figs. 3 and 4. The experimental values 
for crossover time $t^*$, slope and saturation level of `apparent power law' and fit 
of the long-time logarithmic tail provide a sufficient number
of independent measurements for each data set to accuartely determine the two
model parameters, $m$ and $u$. The inset of Fig.5 shows the variation of 
parameter $u$ for the polysiloxane sample 
as function of position on the stress-strain curve, determined by three
different methods: (i) using the difference between $\sigma_{eq}$ and the `apparent power 
law' saturation being $\approx 0.27\, u$, (ii) identifying the numerator of the
long-time inverse-logarithm fit with $u$ and (iii) combining the slope of the 
`apparent power law' $m$ and the crossover time for each plot, $u=(m \, t^*)^{-1/2}$.
The plateau value $u\approx 0.0023$ (and the corresponding plateau for the parameter $m
\approx 700$) strongly depend on the material and temperature, presumably via the 
chain anisotropy, but are also approximately linear functions of the small strain step. 

Considering some experimental ambiguities (such as the effect of non-uniform 
necking) and theoretical simplifications, we consider the proposed model 
to provide a good qualitative description of the data. All three materials 
show the same type of response, with only parameters $m$ and $u$ 
reflecting the difference. Much remains to be done to build a full understanding
of stress relaxation phenomena in nematic elastomers with, it
appears, the unavoidable effects of quenched disorder. One needs to study the effect of 
temperature, in particular, close to the nematic transition point $T_{ni}$. 
The detailed 
role of chain anisotropy $R_\parallel/R_\bot$ remains to be investigated. Correlating 
the stress relaxation with equilibrium autocorrelation functions, expected to 
follow the activated scaling \cite{ising} as the spin-glasses and random nematics do, 
would also be desirable. A refinement of the theoretical model 
is desirable, particularly to address the role of small linear
(semi-soft) corrections to the cubic force. 


We thank H. Finkelmann and R.V. Talroze for the samples used in this study. 
We appreciate discussions with S.F. Edwards, 
M. Warner, S.F. Fridrikh and, in particular, with S.R. Nagel who made the 
analogy with sandpiles. This research has been supported by EPSRC UK.


\end{multicols}

\begin{references}
\bibitem{clark1}T. Bellini, N.A. Clark, C.D. Muzny, L. Wu, C.W. Garland, D.W. Schaefer and 
B.J. Oliver, \PRL \ {\bf 69}, 788 (1992).
\bibitem{garland}L. Wu, B. Zhou, C.W. Garland, T. Bellini and D.W. Schaefer, 
Phys. Rev. E {\bf 51}, 2157 (1995). 
\bibitem{copic}M. \v{C}opi\v{c} and A. Mertelj, \PRL \ {\bf 80}, 1449 (1998).
\bibitem{dozenko}V.S. Dotsenko, {\it Theory of spin glasses and neural networks}, 
World Scientific, Singapore (1994).
\bibitem{ising}D.S. Fisher, \PRL \ {\bf 56}, 416 (1986); \\
\ A.T. Ogielski and D.A. Huse, {\it ibid.} 1298 (1986).
\bibitem{bouchaud}C. Godr\'eche, J.P. Bouchaud and M. Mezard, 
J. Phys. A {\bf 28}, L603 (1995).
\bibitem{goldburg}X.-l. Wu, W.I. Goldburg, M.X. Liu and J.Z. Xue, 
\PRL \ {\bf 69}, 470 (1992).
\bibitem{clark2}T. Bellini, N.A. Clark and D.W. Schaefer, \PRL \ {\bf 74}, 2740 (1995).
\bibitem{review}M. Warner and E.M. Terentjev,  Progr. Polym. Sci. {\bf 21}, 853 (1996).
\bibitem{fridrikh}S.V. Fridrikh and E.M. Terentjev, \PRL \ {\bf 79}, 4661 (1997).
\bibitem{stu}S.M. Clarke, E.M. Terentjev, I. Kundler and H. Finkelmann, 
Macromolecules, (1998) - in press.
\bibitem{nagel}H.M. Jaeger, C.-h. Liu and S.R. Nagel, \PRL \ {\bf 62}, 40 (1989).
\bibitem{weissman}M.B. Weissman, Rev. Mod. Phys. {\bf 60}, 537 (1988).
\bibitem{kupfer}J. K\"upfer and H. Finkelmann, Macromol. Chem. Phys. 
{\bf 195}, 1353 (1994).
\bibitem{mitchell}C.H. Legge, F.J. Davis and G.R.Mitchell, 
J. Phys. II France {\bf 1}, 1253 (1991).
\bibitem{talroze}E.R. Zubarev, T.I. Yuranova, R.V. Talroze, N.A. Plate and 
H. Finkelmann, Macromolecules, (1998) - in press.
\bibitem{foot}{It is important to distinguish between the usual 
structural glass transition, when
the molecular motions in polymeric liquids sieze, 
and the spin-glass-like type of orientational order
\cite{fridrikh} characteristic of randomly disordered nematic systems.}
\bibitem{rubber}{Apart from an ordinary small intra-chain relaxation seen in all 
rubbers, see J.D. Ferry, {\it Viscoelastic properties of polymers}, Wiley, NY (1980). }


\end{references}
\end{document}